\documentclass[twocolumn,showpacs,preprintnumbers,amsmath,amssymb]{revtex4}

\usepackage{graphicx}
\usepackage{dcolumn}
\usepackage{bm}

\begin{document}

\preprint{APS/123-QED}

\title{Droplets displacement and oscillations induced by ultrasonic surface acoustic waves: a quantitative study} 

\author{P. Brunet, M. Baudoin, O. Bou Matar and F. Zoueshtiagh}
\email{philippe.brunet@univ-lille1.fr}
\affiliation{Institut d'Electronique de Micro\'electronique et de Nanotechnologies, UMR CNRS 8520, Cit\'e Scientifique, Avenue Poincar\'e, BP. 60069, 59652 Villeneuve d'Ascq, France}
\date{\today}

\begin{abstract}

We present an experimental study of a droplet interacting with an ultrasonic surface acoustic wave (SAW). Depending on the amplitude of the wave, the drop can either experience an internal flow with its contact-line pinned, or (at higher amplitude) move along the direction of the wave also with internal flow. Both situations appear together with oscillations of the drop free-surface. The physical origins of the internal mixing flow as well as the drop displacement and surface waves are still not well understood. In order to give insights of the underlying physics involved in these phenomena, we carried out an experimental and numerical study. The results suggest that the surface deformation of the drop can be related as a combination between Ôacoustic streamingÕ effect and radiation pressure inside the drop.

\end{abstract}

\pacs{47.55.D- ; *43.25.Nm ; 68.08.Bc ; *43.25.Qp}

\maketitle                              

\section{Introduction}

\subsection{General Issues}

For discrete microfluidics in lab-on-chip devices, it is often necessary to handle small amount of liquids in the form of droplets. These elementary operations are the linear displacement, the splitting or the merging of droplets, or the production of a constant flow inside a drop. At the scale of a micro-litre drop or smaller, the main parameters governing the shape and flow dynamics, are the dynamic viscosity $\mu$ and the surface tension $\sigma$. Non-linear effects originating from inertia are mostly negligible, so that the momentum conservation equation for the fluid reduces to the time-dependent Stokes equation. Consequently, it is hard to induce (chaotic) mixing and continuous re-suspension of particles in a drop as the equation of flow motion is time-reversible. Furthermore, during these operations, the drop is often in contact with ambient atmosphere and evaporation can significantly occur at the typical time-scales of the operations. It is known that when the liquid contains colloidal particles or macromolecules, evaporation leads to a outwards flow in the drop, which in turn leads to a particle clustering in the vicinity of the contact-line\cite{Robert1,Robert2}. The 'coffee rings' originate from this phenomenon, which in practical situation can strongly limit bio-particles detection in lab-on-chip devices (see, e.g., \cite{Nair_Alam}) by forming deposits on DNA microarrays \cite{Blossey}.

\begin{figure}
\begin{center}
(a)\includegraphics[scale=0.25]{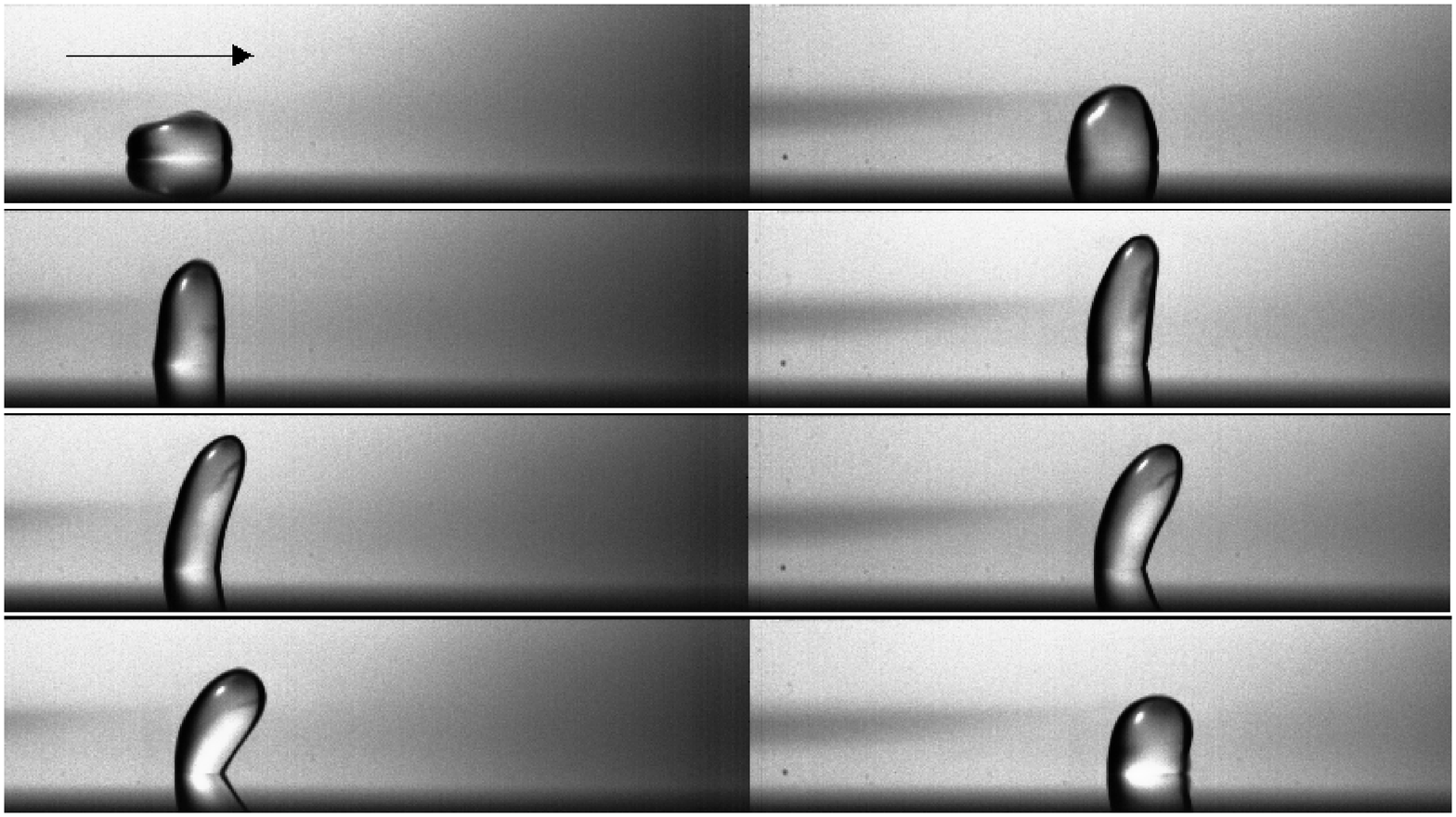}
(b)\includegraphics[scale=2.35]{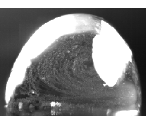}
(c)\includegraphics[scale=0.28]{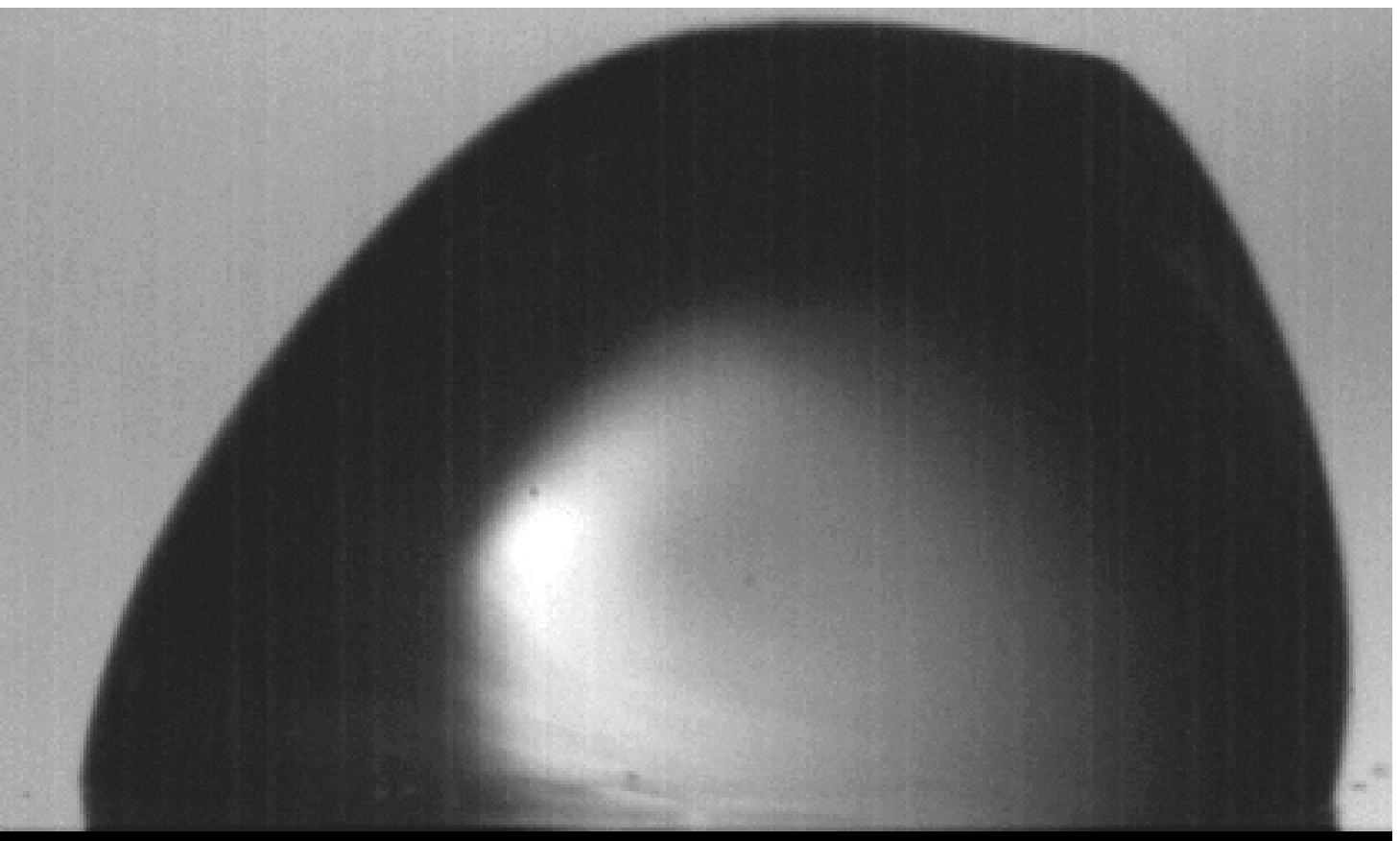}
\caption{(a) Successive snapshots of a water drop displaced by SAW, showing a periodic creeping and jumping flow. The arrow shows the direction of the displacement. (b) Particle tracking inside a steady drop subjected to SAW waves, obtained by time-averaging. (c) Asymmetric shape of a drop moving under the action of SAW (W/G mixture 2), without oscillations. For the three images, the SAW propagates from left to right.}
\label{fig:ptv}
\end{center}
\end{figure}

Another difficulty is to displace such small droplets. As their typical size is smaller than the capillary length $l_c = \sqrt{\sigma / \rho g}$, of the order of 1 to 2 mm for most liquids ($\rho$ being the liquid density), the volume forces are generally overcome by the retention force, $F_R$, acting at the contact-line of the substrate. This force scales with the contact-line perimeter \cite{Dussan}: 

\begin{equation}
F_R = \pi R \sigma (cos(\theta_r) - cos(\theta_a))
\label{eq:retforce}
\end{equation}

Its strength is related to the contact angle hysteresis $\theta_a - \theta_r$, where $\theta_a$ and $\theta_r$ are respectively the advancing and receding contact angles. Real surfaces combining micro- or nano-scale roughness and chemical heterogeneities, can have significant hysteresis \cite{Eggers09}. Furthermore, the aforementioned accumulation of particles near the contact-line increases this hysteresis even more. The coffee-ring effect can be hindered by Marangoni forces if one uses a very volatile solvent \cite{Hu_Larson06}, but with casual liquids and substrates used in biology, volume forces are too weak to re-suspend continuously the particles or to move the drop in a controlled fashion.

One of the challenging issue, is to provoke the continuous unpinning of the contact-line of the drop to both prevent particle accumulation and to obliterate hysteresis. Various authors evidenced a similar unpinning effect with low-frequency mechanical vibrations \cite{Andrieu,DeckerGaroff, Dong_etal06,Noblin04,Brunet07, Long_Chen06}, but it is required that the amplitude of vibrations be of the order of a millimetre to obtain the desired effect. Such vibrations can be unsupported by many fragile surrounding devices. Furthermore, as the pinning forces are often due to micron-sized roughness, it is intuitively more efficient to induce an unpinning flow from waves of shorter wavelength such as those produced by ultrasonic transducers, $\lambda$ ranging from 1 to 300 microns.

\subsection{Qualitative description of the phenomena}

Recently, it has been found that the use of ultrasonic surface acoustic - so-called 'Rayleigh' - waves (SAW) transducers on piezoelectric substrates enables to both agitate inner fluid and to displace a droplet along the direction of the wave propagation \cite{Renaudin,Yeo08,Wixforth}. SAW frequency generally ranges from 5 to 150 MHz, and depending on the operating frequency the dynamics of the droplet can show a host of different behaviours. At low frequencies - about 20 MHz, a 'creeping and jumping' motion of the drop generally occurs, especially for low-viscosity liquids \cite{Renaudin_gfm07}, see Fig.~\ref{fig:ptv}-a. Due to a non-linear coupling between the Rayleigh acoustic wave and the flow, the liquid inside is highly stirred in a rotating flow motion (see Fig.~\ref{fig:ptv}-b), while the interface shows capillary waves. Although several experimental studies demonstrated the reliability of this technique, little is known about the detailed mechanisms that create both the internal flow and the droplet deformations. It is generally admitted that the SAW is radiated inside the droplet at a refraction angle $\phi_r$, analogous to Snellius-Descartes law in optics: 

\begin{equation}
\sin (\phi_r) = \frac{c_l}{c_s}
\end{equation}

\noindent where $c_l$ and $c_s$ denote respectively the speed of sound in the liquid and the solid, and in turn generates acoustic streaming. The angle $\phi_r$ is around 25$^{\circ}$ for water and usual piezo-electric substrates. Regarding droplet dynamics by SAW, a fluid dynamics analysis has been recently attempted \cite{Yeo08}, but results are mostly related to atomisation occurring at high acoustic power. At high frequencies - around 100 MHz, a noticeable contribution on the understanding of drop deformation and displacement, is available in \cite{Schindler} (however at this frequency, no oscillations have been observed). It shows that the flow originates from a carrying force focused in a narrow region inside the drop, starting at the drop edge hit by the wave and directed towards the refraction angle. 

In the general case, this force is partly due to non-linear acoustic streaming \cite{Lighthill,Acousticbook}. The fluid trajectories are described qualitatively by the particle tracking snapshot like that of Fig.~\ref{fig:ptv}-b. In the present frequency range - around 20 MHz, the liquid-air interface is also affected by SAW in three different ways: 

- First, the droplet's left-right symmetry is broken (see Fig.~\ref{fig:ptv}-c) and the asymmetry is more pronounced for larger acoustic power. The asymmetry is responsible for the drop displacement as it makes the front and rear contact-angles becoming respectively larger than $\theta_a$ and smaller than $\theta_r$ \cite{Eggers09}. However, the relative contributions of acoustic streaming and radiation pressure on this asymmetry with respect to the frequency regime, remain unclear. 

- Second, the periodic creeping-jumping motion (see Fig.~\ref{fig:ptv}-a) has a well-defined frequency of the order of 50 to 200 Hz for usual droplet size, and comes with a continuous unpinning of the contact-line. During the deformation cycle, the droplet is strongly stretched and flattened. 

- Finally, besides the global deformation, the droplet surface exhibits small deformations ('trembling') at approximately regular frequency of a few thousands Hertz. However, this is hardly visible on the pictures presented here. 

In any case, the periodic deformations of the interface cannot be attributed to a steady flow. Instead, the acoustic radiation pressure which affects the shape of the liquid-vapour interface, can be invoked for a possible origin of such deformations. Therefore, the periodic motion could be associated with dynamical eigenmodes of the droplet, where capillarity tends to restore a minimal surface energy.

The deep understanding of the involved phenomena is necessary in order to set-up a droplet handling lab-on-chip with minimal acoustic power towards an optimisation of the system. Especially interesting is the use of several combined transducers and short-time pulses, in order to reduce heat (continuous appliance of SAW would cause much heat production) and to protect fragile bio-particles. The current available literature being insufficient for quantitative predictions, the present study aims to carry out quantitative measurements of the displacement and deformation of the drop. We present here such a quantitative study by varying parameters like the droplet volume, the liquid viscosity and the amplitude of the acoustic wave. These results constitute a first step towards the understanding of the detailed coupled acoustic and hydrodynamic mechanisms.

\section{Experimental setup}

\begin{figure}
\begin{center}
\includegraphics[scale=0.375]{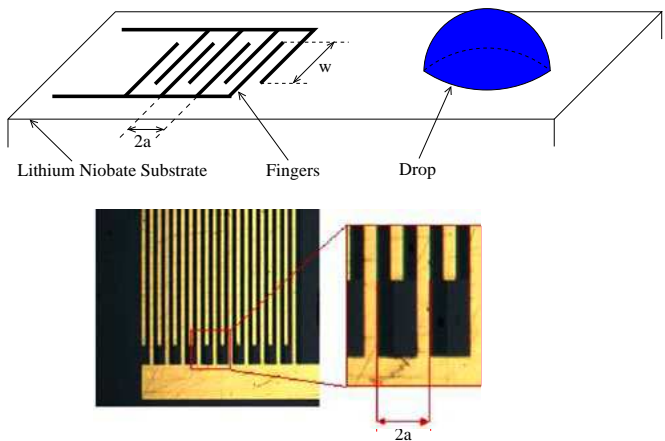}
\caption{Scheme of the SAW interdigitated transducer that actuates with a droplet deposited on the substrate. The wave propagates from left to right. Below: The details of the network of fingers that compose the IDT.}
\label{fig:setup}
\end{center}
\end{figure}

The set-up is depicted in Fig.~\ref{fig:setup}. We used a substrate with piezoelectric properties (Lithium Nyobate, LiNbO$_3$), in order to generate powerful SAW. The transducer is an inter-digitated transducer (IDT) which generates transverse acoustic waves propagating along the surface. The interconnected fingers of the IDT are made of Titane covered by Gold, and designed by a lithographic technique detailed elsewhere \cite{Renaudin}. We applied the periodic sinusoidal voltage with a high-frequency generator (IFR 2023A), amplified with a home-made amplifier. A SAW is generated providing that the voltage frequency is compatible with the space between each track of the IDT. Considering that the space between fingers $a/2$ is 43.75 $\mu$m, giving a wavelength $\lambda$ = $2a$ = 175 $\mu$m, and that the sound velocity in LiNbO$_3$ is 3485 m.s$^{-1}$ for transverse waves, the value of the frequency $f_0$ has to be around 19.5 MHz. In  practice, we found the best actuation to droplets of any size for $f_0$ = 20.375 MHz and we kept this value for all experiments. The properties of the SAW ensure that the amplitude is not attenuated in its direction of propagation, and that it is localised near the surface.

In order to reduce the friction forces, and to study the simpler situation of partial wetting, we treated the surface with hydrophobic coating (monolayer of OTS). The contact angles are: $\theta_a$=105$^{\circ}$ and $\theta_r$=95$^{\circ}$. This weak hysteresis helps to conduct experiments without the additional complexity of pearling at the trailing edge of a moving drop \cite{Podgorski01} that would have occurred otherwise.
The droplet dynamics is acquired with a high-speed camera (Photron SA3) with a zoom lens completed by extension rings. This allows a maximal magnification of about 8 $\mu$m/pixel. The acquisition rate ranged from 2000 to 5000 im/s. To avoid evaporation, special care is devoted to use cold source of light (Schott KL2500) and to acquire pictures no later than a few seconds after the droplet has been put on the substrate.

The liquids used are deionised water and water/glycerol mixtures (W/G), which physical properties at 20$^{\circ}$C are given in Table 1. Glycerol percent weight was varied from 50 to 70, increasing the viscosity by a factor of up to 11, whereas the surface tension stayed close to that of water.

\begin{center}
\begin{table} 
\begin{tabular}{lcccc} 
      Liquid & Kin. viscos. $\nu$\ \ & Surf. tension $\gamma$\ \ & Density $\rho$\\
    & (mm$^2$/s) &   (N/m) &   (g/cm$^3$) \\[3pt] 
       Water &  1.00  & 0.072 & 1.00  \\ 
       W/G. Mix 1 & 5.60  & 0.067 & 1.126  \\
       W/G. Mix 2 & 11.50  & 0.066 & 1.161  \\
\end{tabular}
\caption{Physical properties of liquids}
\label{tab:liq} 
\end{table} 
\end{center}

The measurement of the normal displacement at the surface of the substrate gives the amplitude of the SAW. We used a technique of laser heterodyne interferometer of type Mach-Zender (SH130, B.M. Industries). In brief, the interferometer measures the Doppler shift between a reference beam and a secondary beam modulated at $f_M$ = 70 MHz by an acousto-optic Bragg cell. The secondary beam is reflected onto the surface and hence, is modulated a second time by Doppler effect. The acoustic displacement $d$ is then deduced from these measurements by extracting the relative amplitude between the peaks at $f_M$ and $f_M \pm f_0$ in the power spectrum. The displacement $d$ ranges from 0.3 to 2 nm for the usual experimental conditions.

\section{Results}

\begin{figure}
\begin{center}
\includegraphics[scale=0.12]{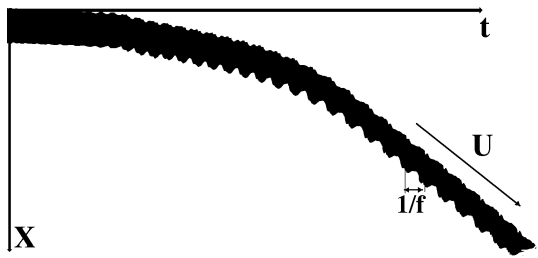}
\caption{Typical spatio-temporal diagram for a moving, creeping and jumping droplet like in Fig.~\ref{fig:ptv}-a.}
\label{fig:starting_phase}
\end{center}
\end{figure}

For each experiment, the dynamics of the drop was extracted using the public domain image processing software ImageJ (\textit{http://rsbweb.nih.gov/ij/}). From a sequence of about 2000 images, we built spatio-temporal diagrams that allowed us to measure both the speed and the frequency of oscillations: we extracted the grey levels along a horizontal line cutting the base of the droplet, and we pilled the lines on top of each others (time axis is horizontal after a rotation of $\pi /2$). A typical motion of drop is pictured in Fig.~\ref{fig:starting_phase} after binary treatment. The total time is about 250 ms. This diagram is then skeletonised to get the position versus time. The velocity $U$ is the slope and the frequency $f$ is extracted from a Fourier transform. It is noticeable that the drop does not start to move or oscillate immediately: it waits about 200 ms before it starts. During this relatively long transient, the oscillations progressively increase, as does the velocity.

\subsection{Droplet displacement}

\begin{figure}
\begin{center}
(a)\includegraphics[scale=0.42]{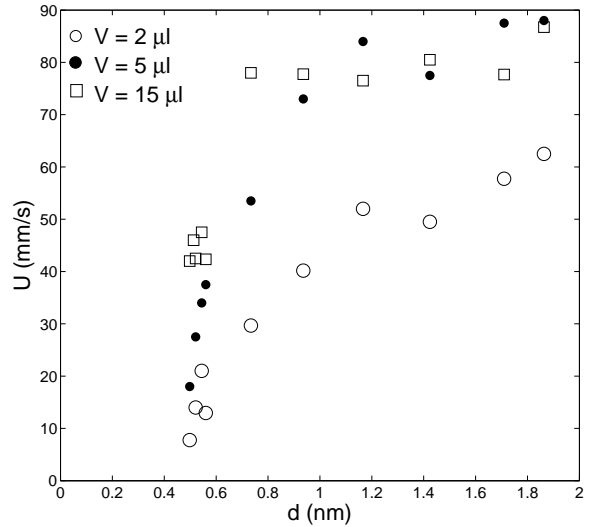}
(b)\includegraphics[scale=0.42]{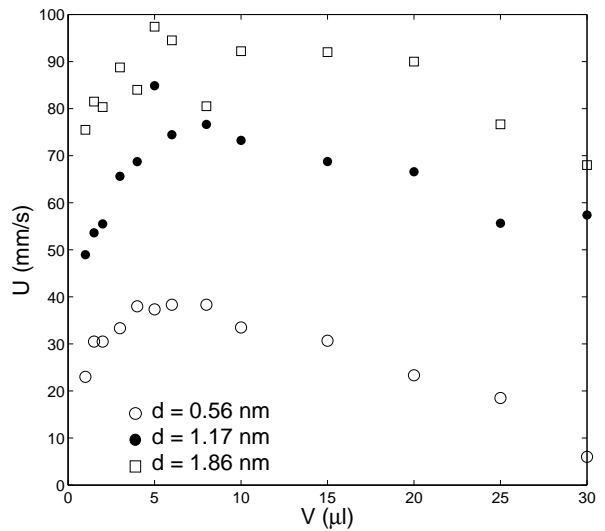}
\caption{(a) Drop velocity versus acoustic displacement $d$, for 3 different volumes (water). (b) Drop velocity versus volume, for 3 values of $d$.}
\label{fig:vitesse}
\end{center}
\end{figure}

We first present the velocity measurements, obtained by varying the acoustic amplitude $d$ at constant volume (Fig.~\ref{fig:vitesse}-a) and by varying the droplet volume $V$ at constant displacement $d$ (Fig.~\ref{fig:vitesse}-b), for water. As stated earlier, the drop displacement results from a balance between driving forces - due to both the streaming and radiation pressure which induce a left/right asymmetry - and dissipation occurring near the contact-line.

\bigskip

\textbf{Influence of acoustic displacement} - As expected, the larger $d$ is, the faster the droplet moves. Also it turns out that $d$ has to be larger than a certain threshold around 0.5 nm in order for the droplet to move. Below this threshold, the droplet stays at the same location but an internal flow motion is observed: this situation is illustrated in Fig.~\ref{fig:ptv}-b. The influence of $d$ is strongly non-linear: after a sharp increase just above threshold, $U$ saturates at a volume-dependent value. 

\bigskip

\textbf{Influence of volume} - Whatever the amplitude of the acoustic wave $d$, the influence of volume shows a maximal velocity at $V$ about 5 $\mu$l (see Fig.~\ref{fig:vitesse}-b). A tentative explanation of this behaviour is developed in the Discussion part. 

\bigskip

\begin{figure}
\begin{center}
\includegraphics[scale=0.29]{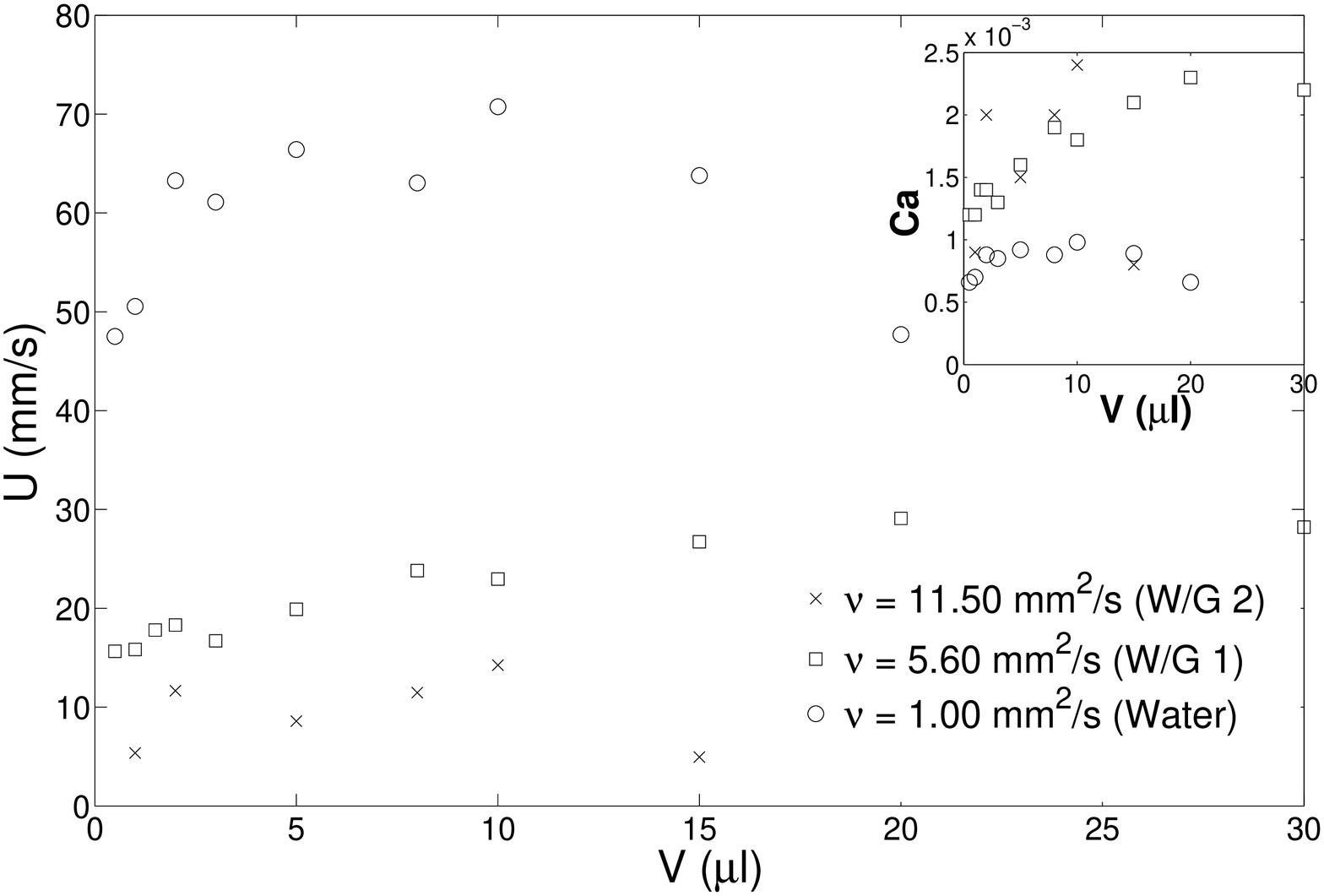}
\caption{Droplet velocity versus volume, for three different values of viscosity. $d$ = 0.99 nm. Inserted: the same measurements with the dimensionless velocity Ca.}
\label{fig:velocity_visco}
\end{center}
\end{figure}

\textbf{Influence of viscosity} - To have a more complete understanding of the coupled acoustic and hydrodynamic effects, we used liquids of various viscosity (see Table 1). Figure \ref{fig:velocity_visco} shows velocity versus volume, for four liquids (Water and W/G Mixtures 1 and 2). The increase of viscosity leads to a decrease of the velocity as expected for moving droplets. Indeed it is known that viscosity is involved in the dissipation near the contact-line \cite{Eggers09} and, hence, an equal driving force displaces viscous drops at lower speed. It is natural to consider the dimensionless velocity, the capillary number 

\begin{equation}
Ca= \frac{\mu U}{\sigma}
\end{equation}

\noindent which is plotted in the insert of Fig.~\ref{fig:velocity_visco}.

It turns out that the data for capillary numbers do not collapse well together, although the order of magnitude is around 1$\times 10^{-3}$ for most measurements. This can be interpreted in the following way: the driving force is due (at least partially) to the acoustic streaming which involves bulk dissipation \cite{Acousticbook}. Hence, streaming effects are promoted by liquid viscosity. The increase of viscosity increases the contact-line dissipation \cite{Eggers09}, but it should also increase the driving force. Therefore, it is consistent that the values of Ca be slightly larger for more viscous liquids.

\subsection{Creeping-jumping oscillatory motion}

\textbf{Influence of viscosity} - First, it is noticeable that oscillations are more and more damped as the viscosity is increased by adding glycerol in the mixture. For instance with mixture 1, oscillations have very small amplitude and are barely measurable, whereas for mixture 2 oscillations completely vanished. Hence, the results of oscillations reported thereafter are for water as this is the only situation where they clearly appear like in Fig.~\ref{fig:ptv}-a. 

\bigskip

\begin{figure}
\begin{center}
\includegraphics[scale=0.35]{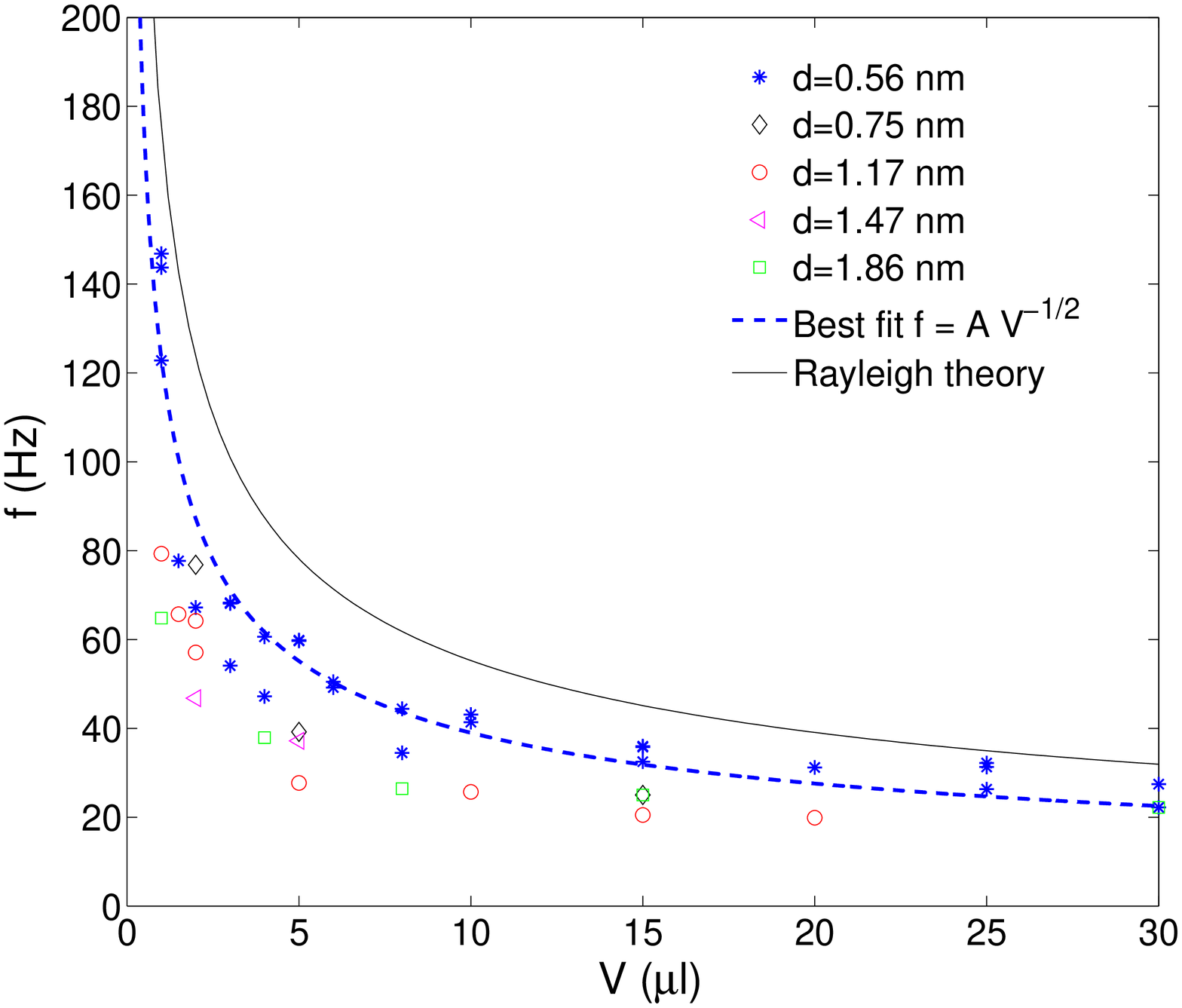}
\caption{Frequency of the creeping-jumping oscillations versus volume, for various acoustic displacement $d$. The dashed blue curve is the best fit $f \sim V^{-1/2}$, and the black plain curve is the prediction by Rayleigh's theory.}
\label{fig:freq}
\end{center}
\end{figure}

\textbf{Frequency} -  In Fig.~\ref{fig:freq}, $f$ is plotted versus volume for different acoustic displacement $d$.

The results clearly show a sharp decrease of $f$ with $V$, in the range $V$ = 0.5 to 5 $\mu$l. A power-law fit leads to an exponent close to one half: $f \sim V^{-1/2}$. This is reminiscent to what is predicted by the theory of Rayleigh-Lamb \cite{Lamb}, for inertial-capillary modes of vibrations in a drop. Rayleigh's seminal calculation is valid for a spherical droplet and for small deformations:

\begin{equation}
f_n = \left( \frac{n (n-1) (n+2)\, \gamma}{3 \pi \rho V} \right)^{1/2}
\label{eq:ray}
\end{equation}

\noindent  in which $f_n$ denotes the resonance frequency of the n$^{th}$ mode of oscillation.

The case n=1 corresponds to pure translation and n=2 corresponds to the lowest mode in which the drop is elongated with an elliptical shape. The morphology of the oscillations suggests a mode n=2 (see Fig. \ref{fig:ptv}-a). To the best of our knowledge, the case of sessile drops has not been treated fully analytically. The pinning force acting at the contact-line strongly constrains the structure of the capillary wave, making its computation more complex. However, Noblin \textit{et al.} \cite{Noblin04} showed that for moderate hysteresis and under strong enough vibrations, the contact-line is constantly unpinned and the constraint due to retention force at the contact-line (eq.\,(\ref{eq:retforce})) is released \cite{Andrieu,DeckerGaroff,Noblin04,Brunet07,Long_Chen06}. This is indeed what is observed in sequences like Fig. \ref{fig:ptv}-a. Hence, we compared out results to what is predicted by  eq.~(\ref{eq:ray}), also plotted in Fig. \ref{fig:freq}. It turns out that Rayleigh-Lamb's theory slightly overestimates the results although the order of magnitude is correctly predicted. This can be interpreted by the fact that SAW tend to induce large amplitude droplet oscillations which are out of the frame of Rayleigh's theory. Intuitively, large amplitude oscillations are of larger period than weak amplitude ones, and this is confirmed by the results of Fig.~\ref{fig:freq}: larger $d$ generates free-surface oscillations of larger amplitude and of smaller frequencies.

\bigskip

\begin{figure}
\begin{center}
\includegraphics[scale=0.45]{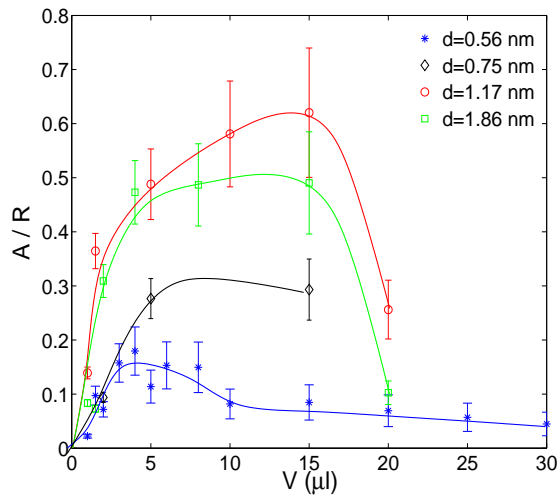}
\caption{Amplitude of the creeping-jumping oscillations of water droplets versus volume, for various acoustic displacements. The curves are guidelines for the eye.}
\label{fig:amploscill}
\end{center}
\end{figure}

\textbf{Amplitude} - In Fig.~\ref{fig:amploscill}, the relative amplitude of oscillations - i.e. the amplitude divided by the mean base radius of the drop - is plotted versus volume, for different acoustic displacements $d$. This amplitude is extracted from spatiotemporal diagrams like that in Fig.~\ref{fig:starting_phase} and corresponds to the maximal lateral deformation of the base of the drop. It turns out that the amplitude is maximal for a $d$-dependent value of volume. These results suggests that oscillations result from a balance between the wave-induced pressure reaching the interface and capillarity, as developed in the Discussion. 

\section{Discussion - Numerical results}

The measurements presented here show several evidences that the dynamics of the drop is due to an interplay between acoustic streaming and radiation pressure. For water droplets at the tested frequency, the SAW induces both a left-right symmetry break-up leading to motion and the creeping-jumping oscillations (the high-frequency 'trembling' oscillations, mentioned earlier, will be treated in a further study). It is still unclear what the respective weight of streaming and radiation pressure are on the drop  left/right asymmetry. Schindler \textit{et al.} \cite{Schindler} showed that at high frequencies, acoustic streaming can lead to a conservative force potential in the drop towards the direction of propagation, headed to the top and inducing a deformation like in Fig.~\ref{fig:ptv}-c. However, the streaming flow cannot explain the creeping-jumping oscillations in which the free-surface of the drop is pushed upwards.

The acoustic streaming inside the droplet comes from dissipation due to liquid shear ($\mu$) and bulk ($\mu_b$) viscosities. The internal streaming flow is more intense for larger viscosities and frequency \cite{Acousticbook}. On the contrary, the radiation pressure is directly proportional to the energy which reaches the free-surface \cite{Acousticbook}. Consequently, the larger the viscous dissipation is, the smaller this energy is. The relative importance of both effects is strongly dependent on the structure of the acoustic wave diffracted in the drop, and on how much the wave is attenuated. 

It is relevant to compare the typical drop size with the length of attenuation of the acoustic wave which, in a fluid, is given by \cite{Acousticbook}:

\begin{equation}
L = \frac{2 \rho c_l^3}{(2 \pi f_0)^2} \left(\frac{1}{\frac{4}{3} \mu + \mu_b} \right)
\label{eq:atten}
\end{equation}

\begin{figure}
\begin{center}
\includegraphics[scale=0.45]{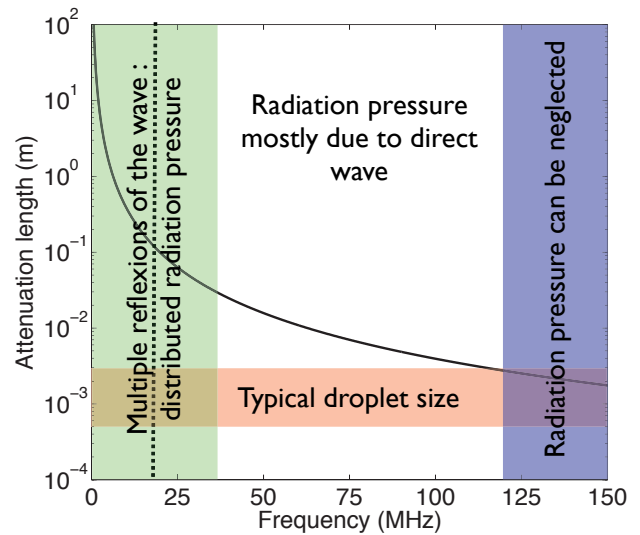}
\caption{The different regimes of the relative influence of radiation pressure versus frequency. The plain curve stands for the attenuation length $L$ in pure water ($c_l$=1476 m/s and $\mu_b$=2.8$\times 10^{-3}$ Pa.s) given by eq.~(\ref{eq:atten}).}
\label{fig:att}
\end{center}
\end{figure}

This length is plotted versus $f_0$ in Fig.~\ref{fig:att}, with respect to the drop size. The length $L$ is equal or lower than the size of droplets for a frequency of about 125 MHz or above (see Fig.~\ref{fig:att}) and defines a range of frequency where radiation pressure is negligible. This is far beyond the values of $f_0$ used here and, therefore, both streaming and radiation pressure are effective in our study.

\begin{figure}
\begin{center}
(a)\includegraphics[scale=3.5]{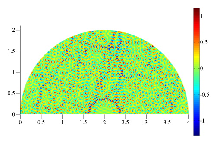}
(b)\includegraphics[scale=3.5]{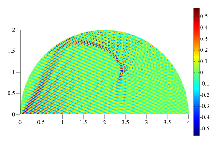}
(c)\includegraphics[scale=3.5]{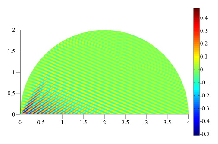}
\caption{Numerical simulations of a SAW inside an non-deformable half-spherical drop, $f_0$=20.375 MHz and $d$= 2 nm. Colour levels stand for pressure. (a) Weak attenuation in the case of water. (b) Intermediate attenuation, 25 times larger than (a). (c) Strong attenuation, 100 times larger than (a).}
\label{fig:simus}
\end{center}
\end{figure}

\begin{figure}
\begin{center}
\includegraphics[scale=3.8]{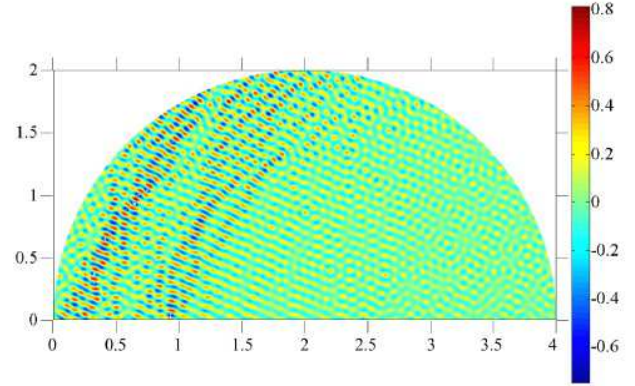}
\caption{Pressure due to the coherent direct wave in the situation of Fig. \ref{fig:simus}-a, with chaotic multiple reflections. Obtained by averaging over several runs, see text for details.}
\label{fig:simu_moy}
\end{center}
\end{figure}

To confirm this trend, at least qualitatively, we carried out numerical simulations with a finite-elements software (COMSOL Multiphysics 3.4). The model uses a drop of half-spherical fixed shape (contact angle = 90 $^{\circ}$, radius = 2 mm) subjected to a SAW. The General Form Modes of COMSOL Multiphysics with two domains has been used, one for the piezoelectric substrate and one for the water droplet. In the water droplet, the linear equation of acoustic in a thermo-viscous fluid has been considered. The boundary conditions used at the drop/ambient air interface are stress-free conditions. As we are interested in the linear radiation of the SAW propagating at the substrate surface into the water, the continuity of the normal stresses and displacements has been assumed between the piezoelectric substrate and the liquid \cite{Campbell}. This last point is justified by the fact that in our experiences the amplitude of SAW is always small in comparison to the acoustic wavelength.

Figure \ref{fig:simus} shows the pressure distribution (a) in a water drop for $f_0$=20 MHz and in liquids where the attenuation is (b) 25 times larger and (c) 100 times larger, for the same $f_0$. When weakly attenuated, as in water for a frequency of 20 MHz (Fig.~\ref{fig:simus}-a), the wave structure is generally chaotic due to multiple reflections as the drop acts like a chaotic cavity. If the radius is slightly modified, the wave structure is completely different except for the coherent part in the direct wave. To extract this coherent part we computed the wave structure for 15 droplets of radius ranging from 2 to 2.35 mm, and we averaged the contribution of each case. This calculation procedure, where stationary wave pattern are determined in non-deformable drops of increasing size, is justified by the fact that the time required for the wave to propagate across the 2 mm drop is about 2 $\mu$s, which is much lower than the observed characteristic deformation time of the drop.

The resulting data exhibits the coherent part of the wave, as shown in Fig.\ref{fig:simu_moy}. In this case, the wave reaches the free-surface whereas in the two other cases (Fig.~\ref{fig:simus}-b and c) it is strongly attenuated and hardly reaches the free-surface. It is clear that radiation pressure will only operate in the first case. The second case is streaming-dominated and should not show global free-surface oscillations. The action of radiation pressure is reminiscent to another study \cite{Bastien02}, although for much smaller frequencies and a different set-up, where the free-surface of a liquid layer is deformed by acoustic vibrations of a plate.

The coherent part of the wave is highly left-right asymmetric and the associated radiation pressure therefore contributes to the asymmetry of the droplet. On the contrary the chaotic part of the wave induces a radiation pressure homogeneously distributed on the surface of the wave. Therefore, only the direct coherent wave contributes to the displacement of the droplet whereas the chaotic part should be the driving effect of the free-surface oscillations. The computations showed that the chaotic part vanishes for smaller $f_0$ than the coherent direct part. Indeed, for the existence of multiple reflections it is required that the length $L$ be several times larger than the drop size. Therefore, we defined two other regimes: a 'multiple reflections' domain at lower $f_0$, and an intermediate regime where only the coherent part of the direct wave contributes to the radiation pressure.

The use of liquids more viscous than water have similar effects on the relative contribution of radiation pressure as the prescription of a higher frequency: it decreases $L$. In the cases displayed in Fig.~\ref{fig:simus}-b and Fig.~\ref{fig:simus}-c, the corresponding frequencies would be 100 MHz and 200 MHz, respectively. The use of W/G mixtures increases both $\mu$ and $\mu_b$. The ratio between $\mu_b$ and $\mu$ is 2.8 for water, and it should be quite close to this value for W/G mixtures of low viscosity like ours. It is sensible to interpret the absence of oscillations for W/G droplets by a decrease of the length $L$: as the wave is more strongly attenuated when arriving at the droplet free-surface, the radiation pressure is weaker than for water drops. In Fig.~\ref{fig:att}, the curve of $L$ vs.~$f_0$ for W/G mixtures would be just translated downwards.

To observe the creeping-jumping oscillations, the radiation pressure has to be strong enough to cause the unpinning of the droplet contact-line (otherwise, as the drop volume is constant, the height of the drop cannot vary significantly). Our measurements and computations suggest that the chaotic part of the wave is required to obtain such pressure: this condition is fulfilled only for frequencies included in the lowest range domain and for low-viscous liquids. This is to be related to the results of Fig.~\ref{fig:freq} for the selection of the frequency: once the radiation pressure pushes the top of the drop and shrinks its base surface, the contribution of this pressure decreases as the height of the drop increases. Therefore, the drop retracts under the action of capillary forces, leading to a frequency comparable to that of the first eigenmode of the drop driven by an inertial-capillary balance. The maxima of amplitude found for a $d$-dependent intermediate volume (Fig. \ref{fig:amploscill}) is consistent with this interpretation: surface tension hinders large oscillations for small drops, as capillary pressure, scaling with interface curvature, is bigger; whereas large drops are subjected to lesser radiation pressure because the acoustic wave is more damped before reaching the interface. 

Finally, these calculations suggest a possible explanation for the maximal drop velocity for water, obtained for intermediate drop volume (see Fig.~\ref{fig:vitesse}). If the drop is too small, only a part of the SAW is transferred to the droplet whereas a significant part of the acoustic energy goes on propagating in the solid after the drop \cite{Jiao}. On the contrary, drops of larger size collect most of the acoustic energy for their displacement and deformation. Bigger drops collect about the same amount of energy as intermediate drops, but they will be slowed down simply because they have more inertia. The drop optimal size is dictated by the attenuation length in the solid, in the X direction. This attenuation length is no more linked to the thermo-viscous absorption in the liquid, but to the imaginary part $k_i$ of the Rayleigh wave wavenumber, which is of leaky type when propagating in a substrate in contact with a liquid. For a Lithium Niobate substrate loaded by water $k_i \simeq$ 2.3 10-5 $\times f_0$ Np/m \cite{Campbell,Shiokawa}, more than two orders of magnitude higher than the attenuation coefficient at $f_0$=20 MHz. This corresponds to the fact that the Rayleigh wave efficiently looses its energy by radiating in the liquid. The extraction of the amplitude at the solid-liquid interface gives an attenuation length of about 2.13 mm, corresponding to a drop volume of about 5 $\mu$l. This may explain the maxima of velocity observed in Fig.~\ref{fig:vitesse}.

\section{Conclusion}

We have presented here a quantitative study of droplet dynamics, when actuated by a SAW. Most of previous studies of SAW ignored the effect of radiation pressure as they were carried out at higher frequencies. In our range of frequency and liquid viscosity, a combination between streaming and radiation pressure effects was observed and their respective contributions on the observed dynamics was studied. The free-surface oscillations are mostly due to radiation pressure, whereas internal flow is due to streaming. This is evidenced by both experiments at different viscosity, and computations. Contributions of both phenomena are of interest, for an optimal micro-fluidics mixing efficiency. Future studies will focus on the influence of frequency in order to determine the relative weight of both mechanisms in the drop's asymmetry.



\end{document}